\documentclass[aps,prl,preprint,tightenlines,superscriptaddress,showpacs,byrevtex]{revtex4}
\usepackage{graphicx} 
\usepackage{dcolumn}  
\usepackage{multirow}
\usepackage{hhline}
\usepackage{color}      

\graphicspath{{ps}}

\begin{document}


\preprint{\vbox{ \hbox{   }
						\hbox{Belle Preprint{\it 2020-21}}
						\hbox{KEK Preprint{\it 2020-38}}
						\hbox{University of Cincinnati Preprint{\it UCHEP-21-05}}
                        \hbox{  \includegraphics[width=2.5cm]{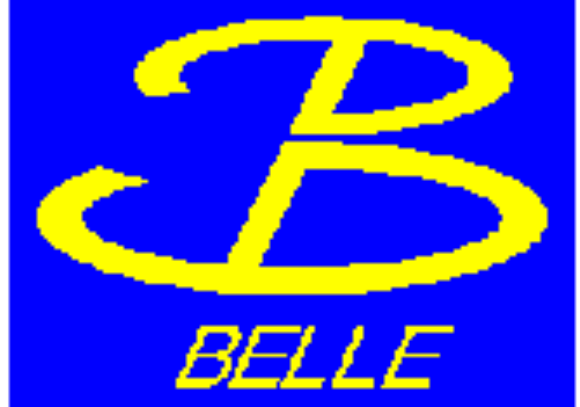}
}
}}

\title{ \quad\\[1.8cm] Measurement of ${\mathcal B}$($B_s \rightarrow D_s X$) with $B_s$
Semileptonic Tagging}

\noaffiliation
\affiliation{University of the Basque Country UPV/EHU, 48080 Bilbao}
\affiliation{Beihang University, Beijing 100191}
\affiliation{Brookhaven National Laboratory, Upton, New York 11973}
\affiliation{Budker Institute of Nuclear Physics SB RAS, Novosibirsk 630090}
\affiliation{Faculty of Mathematics and Physics, Charles University, 121 16 Prague}
\affiliation{Chonnam National University, Kwangju 660-701}
\affiliation{University of Cincinnati, Cincinnati, Ohio 45221}
\affiliation{Deutsches Elektronen--Synchrotron, 22607 Hamburg}
\affiliation{Key Laboratory of Nuclear Physics and Ion-beam Application (MOE) and Institute of Modern Physics, Fudan University, Shanghai 200443}
\affiliation{Justus-Liebig-Universit\"at Gie\ss{}en, 35392 Gie\ss{}en}
\affiliation{II. Physikalisches Institut, Georg-August-Universit\"at G\"ottingen, 37073 G\"ottingen}
\affiliation{SOKENDAI (The Graduate University for Advanced Studies), Hayama 240-0193}
\affiliation{Gyeongsang National University, Chinju 660-701}
\affiliation{Hanyang University, Seoul 133-791}
\affiliation{University of Hawaii, Honolulu, Hawaii 96822}
\affiliation{High Energy Accelerator Research Organization (KEK), Tsukuba 305-0801}
\affiliation{J-PARC Branch, KEK Theory Center, High Energy Accelerator Research Organization (KEK), Tsukuba 305-0801}
\affiliation{Forschungszentrum J\"{u}lich, 52425 J\"{u}lich}
\affiliation{IKERBASQUE, Basque Foundation for Science, 48013 Bilbao}
\affiliation{Indian Institute of Science Education and Research Mohali, SAS Nagar, 140306}
\affiliation{Indian Institute of Technology Bhubaneswar, Satya Nagar 751007}
\affiliation{Indian Institute of Technology Guwahati, Assam 781039}
\affiliation{Indian Institute of Technology Hyderabad, Telangana 502285}
\affiliation{Indian Institute of Technology Madras, Chennai 600036}
\affiliation{Institute of High Energy Physics, Chinese Academy of Sciences, Beijing 100049}
\affiliation{Institute of High Energy Physics, Vienna 1050}
\affiliation{Institute for High Energy Physics, Protvino 142281}
\affiliation{INFN - Sezione di Napoli, 80126 Napoli}
\affiliation{INFN - Sezione di Torino, 10125 Torino}
\affiliation{Advanced Science Research Center, Japan Atomic Energy Agency, Naka 319-1195}
\affiliation{J. Stefan Institute, 1000 Ljubljana}
\affiliation{Institut f\"ur Experimentelle Teilchenphysik, Karlsruher Institut f\"ur Technologie, 76131 Karlsruhe}
\affiliation{Kennesaw State University, Kennesaw, Georgia 30144}
\affiliation{King Abdulaziz City for Science and Technology, Riyadh 11442}
\affiliation{Korea Institute of Science and Technology Information, Daejeon 305-806}
\affiliation{Korea University, Seoul 136-713}
\affiliation{Kyungpook National University, Daegu 702-701}
\affiliation{LAL, Univ. Paris-Sud, CNRS/IN2P3, Universit\'{e} Paris-Saclay, Orsay}
\affiliation{\'Ecole Polytechnique F\'ed\'erale de Lausanne (EPFL), Lausanne 1015}
\affiliation{P.N. Lebedev Physical Institute of the Russian Academy of Sciences, Moscow 119991}
\affiliation{Faculty of Mathematics and Physics, University of Ljubljana, 1000 Ljubljana}
\affiliation{Ludwig Maximilians University, 80539 Munich}
\affiliation{Luther College, Decorah, Iowa 52101}
\affiliation{University of Maribor, 2000 Maribor}
\affiliation{Max-Planck-Institut f\"ur Physik, 80805 M\"unchen}
\affiliation{School of Physics, University of Melbourne, Victoria 3010}
\affiliation{University of Mississippi, University, Mississippi 38677}
\affiliation{University of Miyazaki, Miyazaki 889-2192}
\affiliation{Moscow Institute of Physics and Technology, Moscow Region 141700}
\affiliation{Graduate School of Science, Nagoya University, Nagoya 464-8602}
\affiliation{Universit\`{a} di Napoli Federico II, 80055 Napoli}
\affiliation{Nara Women's University, Nara 630-8506}
\affiliation{National Central University, Chung-li 32054}
\affiliation{National United University, Miao Li 36003}
\affiliation{Department of Physics, National Taiwan University, Taipei 10617}
\affiliation{H. Niewodniczanski Institute of Nuclear Physics, Krakow 31-342}
\affiliation{Nippon Dental University, Niigata 951-8580}
\affiliation{Niigata University, Niigata 950-2181}
\affiliation{Novosibirsk State University, Novosibirsk 630090}
\affiliation{Osaka City University, Osaka 558-8585}
\affiliation{Pacific Northwest National Laboratory, Richland, Washington 99352}
\affiliation{Panjab University, Chandigarh 160014}
\affiliation{Peking University, Beijing 100871}
\affiliation{University of Pittsburgh, Pittsburgh, Pennsylvania 15260}
\affiliation{Theoretical Research Division, Nishina Center, RIKEN, Saitama 351-0198}
\affiliation{University of Science and Technology of China, Hefei 230026}
\affiliation{Seoul National University, Seoul 151-742}
\affiliation{Showa Pharmaceutical University, Tokyo 194-8543}
\affiliation{Soongsil University, Seoul 156-743}
\affiliation{Sungkyunkwan University, Suwon 440-746}
\affiliation{School of Physics, University of Sydney, New South Wales 2006}
\affiliation{Department of Physics, Faculty of Science, University of Tabuk, Tabuk 71451}
\affiliation{Tata Institute of Fundamental Research, Mumbai 400005}
\affiliation{Toho University, Funabashi 274-8510}
\affiliation{Department of Physics, Tohoku University, Sendai 980-8578}
\affiliation{Earthquake Research Institute, University of Tokyo, Tokyo 113-0032}
\affiliation{Department of Physics, University of Tokyo, Tokyo 113-0033}
\affiliation{Tokyo Metropolitan University, Tokyo 192-0397}
\affiliation{Virginia Polytechnic Institute and State University, Blacksburg, Virginia 24061}
\affiliation{Wayne State University, Detroit, Michigan 48202}
\affiliation{Yamagata University, Yamagata 990-8560}
\affiliation{Yonsei University, Seoul 120-749}
  \author{B.~Wang}\affiliation{University of Cincinnati, Cincinnati, Ohio 45221}\affiliation{Max-Planck-Institut f\"ur Physik, 80805 M\"unchen}  
  \author{K.~Kinoshita}\affiliation{University of Cincinnati, Cincinnati, Ohio 45221} 
  \author{H.~Aihara}\affiliation{Department of Physics, University of Tokyo, Tokyo 113-0033} 
  \author{D.~M.~Asner}\affiliation{Brookhaven National Laboratory, Upton, New York 11973} 
  \author{T.~Aushev}\affiliation{Moscow Institute of Physics and Technology, Moscow Region 141700} 
  \author{R.~Ayad}\affiliation{Department of Physics, Faculty of Science, University of Tabuk, Tabuk 71451} 
  \author{V.~Babu}\affiliation{Deutsches Elektronen--Synchrotron, 22607 Hamburg} 
  \author{I.~Badhrees}\affiliation{Department of Physics, Faculty of Science, University of Tabuk, Tabuk 71451}\affiliation{King Abdulaziz City for Science and Technology, Riyadh 11442} 
  \author{A.~M.~Bakich}\affiliation{School of Physics, University of Sydney, New South Wales 2006} 
  \author{P.~Behera}\affiliation{Indian Institute of Technology Madras, Chennai 600036} 
% ...
\collaboration{The Belle Collaboration}

\begin{abstract}
We report the first direct measurement of the inclusive branching fraction ${\mathcal B}(B_s \rightarrow D_s X)$ via $B_s$ tagging in $e^+e^-\to\Upsilon$(5S) events.
Tagging is accomplished through a partial reconstruction of semileptonic decays $B_s \rightarrow D_s X \ell \nu$, where $X$ denotes unreconstructed additional hadrons or photons and $\ell$ is an electron or muon. 
With 121.4 fb$^{-1}$ of data collected at the $\Upsilon$(5S) resonance by the Belle detector at the KEKB asymmetric-energy $e^+ e^-$ collider,  we obtain ${\mathcal B}(B_s \rightarrow D_s X)$ = 
$(60.2 \pm 5.8 \pm 2.3)$\%, where the first uncertainty is statistical and the second is systematic.
\end{abstract}


\maketitle

\tighten

{\renewcommand{\thefootnote}{\fnsymbol{footnote}}}
\setcounter{footnote}{0}


The study of $B_s$-meson properties at the $\Upsilon$(5S) resonance may provide important insights
into the CKM matrix and hadronic structure, as well as sensitivity to new
physics phenomena~\cite{falk2000, petrak2001, atwood2002}. 
The branching fraction for the inclusive decay $B_s \rightarrow D_s X$ plays an important role in the determination of the $B_s$ production rate in $\Upsilon$(5S) events\cite{nocharge}.
This rate, usually expressed as the fraction $f_s$ of $b\bar b$ events at the $\Upsilon$(5S), is necessary for measuring absolute rates and branching fractions.
Two experiments at LEP, ALEPH~\cite{aleph-result} and OPAL~\cite{opal-result},  measured the product branching fraction ${\mathcal B}(\overline{b}~\rightarrow B_s^0)~\cdot {\mathcal B}(B_s^0 \rightarrow D_s X)$.
The branching fraction ${\mathcal B}(B_s^0~\rightarrow~D_s X)$ was evaluated using a model-dependent value of ${\mathcal B}(\overline{b}~\rightarrow~B_s^0)$ and was subject to large statistical and theory uncertainties. 
Belle measured the branching fractions of $\Upsilon(5{\rm S}) \rightarrow D_s X$ and $\Upsilon(5{\rm S}) \rightarrow D^0 X$~\cite{belle-result} with 1.86~fb$^{-1}$ of data collected at the $\Upsilon$(5S) energy. 
These are related to the inclusive $B_s$ branching fractions to $D_s$ and $D^0/\bar D^0$ by the following relations,
\begin{eqnarray}
\label{eq:Bs2Dx}
  {\mathcal B}(\Upsilon(5{\rm S}) \rightarrow D_x X)/2 &= & f_s \cdot {\mathcal B}(B_s
  \rightarrow D_x X)\nonumber \\ 
  &&+ f_{q} \cdot {\mathcal B}(B \rightarrow D_x X),
\end{eqnarray}
where $D_x$ is $D_s$ or $D^0/\bar D^0$,  $f_s$ is the fraction of $\Upsilon$(5S) events containing  $B_s$-meson pairs, and $f_{q}$ is the fraction containing charged or neutral $B$  pairs.
Using the measured value of ${\mathcal B}(\Upsilon(5{\rm S}) \rightarrow D^0 X)$\cite{belle-result}, and assuming $f_q=1-f_s$ and ${\mathcal B}(B_s \rightarrow D^0 X +c.c.)=8\pm 7\%$\cite{CLEOBs}, which was estimated based on phenomenological arguments, Belle found $f_s=(18.1\pm 3.6\pm 7.5)\%$\cite{belle-result}.
This input, with the measured ${\mathcal B}(\Upsilon(5{\rm S}) \rightarrow D_s X)$\cite{belle-result}, was used to evaluate ${\mathcal B}(B_s \rightarrow D_s X)=(91\pm 18\pm 41)\%$\cite{belle-result}.
The current world average, $(93\pm 25)$\%\cite{pdg2020}, is based on measurements made with the methods described above, which rely on model-dependent assumptions.

In this paper, we present the first direct measurement of ${\mathcal B}(B_s~\rightarrow~D_s X)$ using a $B_s$ semileptonic tagging method with $\Upsilon$(5S) events. 
Throughout this paper, the inclusive branching fraction ${\mathcal B}(B_s \rightarrow D_s X$) is defined as the mean number of $D_s$-mesons per $B_s$ decay.

We use a data sample of 121.4 fb$^{-1}$, collected with the Belle detector~\cite{Belle} at the KEKB asymmetric-energy $e^+e^-$ collider~\cite{kekb} operating near the $\Upsilon$(5S) resonance. The Belle detector is a general-purpose large-solid-angle spectrometer consisting of a silicon vertex detector (SVD), a central drift chamber (CDC), an array of aerogel threshold Cherenkov counters (ACC), a barrel-like arrangement of time-of-flight scintillation counters (TOF), and an electromagnetic calorimeter (ECL) located inside a superconducting solenoid coil that provides a 1.5 T magnetic field. Outside the coil, an iron flux-return yoke is instrumented to detect $K_L^0$-mesons and to identify muons (KLM). A detailed description of the detector can be found in Ref.~\cite{Belle}.


All charged tracks, except those from $K_S^0$ decay, are required to be consistent with originating from the interaction point (IP), with the point of closest approach to the IP within 2.0~cm along the beam axis and within 0.5~cm in the plane transverse to the beam. 
Additionally, all tracks must have, within the SVD, at least one associated hit in the plane transverse to the beam and two hits along the beam axis. 
To suppress the continuum background from $e^+e^- \rightarrow q \overline{q}$ with $q=u,\ d,\ s,$ or $c$, we require that the variable $R_2$, the ratio of second- to zeroth-order Fox-Wolfram moments~\cite{wolfram}, be less than $0.4$.
Kaon and pion hypotheses are assigned to the tracks based on  likelihood, which is calculated using information from the Cherenkov light yield in the ACC, the time-of-flight information of the TOF, and the specific ionization ($dE/dx$) in the CDC.
Charged kaon (pion) candidates are required to have a kaon/pion likelihood ratio
${\cal L}_K/({\cal L}_K+{\cal L}_\pi)>0.6\ (<0.6)$. 
The angle between each lepton and the positron beam is required to be between 18$^{\circ}$ and 150$^{\circ}$ for electrons and between 25$^{\circ}$ and 145$^{\circ}$ for muons. 
Selected electrons and muons must have a minimum momentum of 1.0~GeV/$c$ in the $e^+e^-$ center-of-mass (CM) frame. 
An electron/pion likelihood ratio (${\cal L}_e$) is calculated based on information from the CDC, ACC, and ECL.
A muon/hadron likelihood ratio is calculated based on information from the KLM.
Tracks with ${\cal L}_{e}>0.8$ (${\cal L}_{\mu}>0.8$) are included as electrons (muons) in the analysis. 
The efficiency for electron (muon) tracks to pass this criterion is ($94.7\pm 0.2$)\% (($96.7\pm 0.2$)\%).

The neutral intermediate particles $\phi, K_S^0$ and $K^{*0}$\cite{chargeconjugate} are reconstructed from charged tracks. 
For $\phi\to K^+K^-$ reconstruction, any pair of oppositely charged kaons with invariant mass within 15~MeV/$c^2$ of the $\phi$ nominal mass\cite{pdg2020} is considered to be a $\phi$ candidate. 
The $K_S^0$ candidates are reconstructed via the decay $K_S^0\to \pi^+\pi^-$, following standard criteria \cite{KScitation}, and are further required to have an invariant mass within 20~MeV/$c^2$ ($\approx$4.4~$\sigma$ in resolution) of the nominal mass.
For $K^{*0}\to K^+\pi^-$, the candidate tracks are oppositely charged $K$ and $\pi$, with invariant mass within 50~MeV/$c^2$. 

Candidates for $D_s^+$ are reconstructed in the final states $\phi\pi^+$, $K_S^0K^+$, and $\bar K^{*0}K^+$.
The CM momentum of the candidate is required to be in the range 0.5 GeV/$c$ -- 3.0 GeV/$c$. 
Candidates with invariant mass in the range 1.92-2.02~GeV/$c^2$ are considered. 
For $\phi \pi^+$ and $\bar K^{*0} K^+$ modes, a vertex fit is performed for the three tracks used to reconstruct the candidate, and the $\chi^2$ of the fit output is required to be less than 100.
Nearly all correctly reconstructed $D_s$, ($98.1\pm 0.1)$\%, are found to pass this requirement.
The decays $D_s^+ \rightarrow \phi (K^+K^-) \pi^+$ and $D_s^+ \rightarrow \bar K^ {*0} (K^- \pi^+) K^+$ are transitions of a pseudoscalar particle to a vector and a pseudoscalar, with the vector decaying to two pseudoscalars. 
To suppress combinatorial background, we require $|\cos\theta_{\rm hel}| > 0.5$, where the helicity angle $\theta_{\rm hel}$ is defined as the angle between the momentum of the $D_s^+$ and $K^+$ ($\pi^+$) in the rest frame of the $\phi$ ($\bar K^{*0}$) resonance.

We tag $B_s$ events through  a ``partial reconstruction'' of the semileptonic decay $B_s^0 \rightarrow D_s^- X \ell^+ \nu$, with the $D_s^-$ modes $\phi\pi^-$ and $K_S^0K^-$, using a procedure similar to one applied at the $\Upsilon(4S)$ resonance~\cite{CLEO1989}, where a lepton (electron or muon) is paired with a charm meson to form a $B$ candidate. 
In contrast to the $\Upsilon$(4S), where the exclusive production of $B\bar B$ ensures that each $B$-meson's total energy is half the CM energy, $\sqrt{s}$/2, the $B_s$'s in $\Upsilon$(5S) events occur predominantly in $B_s^*\bar B_s^*$ events.
In this case the energy of each $B_s$ is well approximated as $\sqrt{s}/2-\delta E$, where $\delta E/c^2$ is the $B_s^*-B_s$ mass difference. 
We use $\delta E$=47.3~MeV.
We thus define the ``missing mass squared'' of the selected $D_s^-\ell^+$ candidate as
\begin{equation}
  \label{eq:miss-mass}
  M_{\rm miss}^2 = (\sqrt{s}/2 - \delta E - E_{D\ell}^*)^2 - (p_{D\ell}^*)^2,
\end{equation}
where $E_{D\ell}^*$ and $p_{D\ell}^*$ are the energy and momentum of the $D_s\ell$ system in the CM frame. 
The distribution in $M_{\rm miss}^2$ for tagged $B_s$ represents the undetected neutrino plus additional low-momentum daughters of excited $D_s$, photons and pions, and is expected to peak broadly at $M_{\rm miss}^2=0$.
The thrust angle, $\theta_{\rm thrust}$ is defined as the angle between the thrust axis\cite{thrust} of the selected $D_s\ell$ system and that of the remaining tracks in the event. 
To suppress continuum background, we require $|\cos\theta_{\rm thrust}| < 0.8$.
In events with more than one tag candidate, we perform a combined fit on each candidate's three-track $D_s$ vertex, and on the vertex of the extrapolated $D_s$ trajectory with the lepton, and select the candidate having the smallest $\chi^2$.

\begin{table}[tb]

  \centering
  \caption{Signal-side $D_s$ reconstruction efficiencies, by tag-side and signal-side $D_s$ decay channel.}
  \label{tab:eff-sig-ds}

  \begin{tabular}{l l l }
    \hline \hline
    Tag Channel & Signal Channel & Efficiency (\%) \\
    \hline 
    \multirow{3}{*}{$\phi  \pi$} & $\phi\{K^+K^-\} \pi$ & 26.1 $\pm$ 0.5 \\
    & $K_S^0\{\pi^+\pi^-\} K$ & 38.5 $\pm$ 0.6 \\
    & $K^{*0}\{K^\pm\pi^\mp\} K$ & 24.6 $\pm$ 0.5 \\
    \hline
    \multirow{3}{*}{$K_S^0 K$} & $\phi\{K^+K^-\} \pi$ & 27.6 $\pm$ 0.5 \\
    & $K_S^0\{\pi^+\pi^-\} K$ & 37.8 $\pm$ 0.6 \\
    & $K^{*0}\{K^\pm\pi^\mp\}  K$ & 24.6 $\pm$ 0.4 \\
    \hline \hline
  \end{tabular}

\end{table}

\begin{figure}[tb] 
  \centering
  \includegraphics[width=7cm]{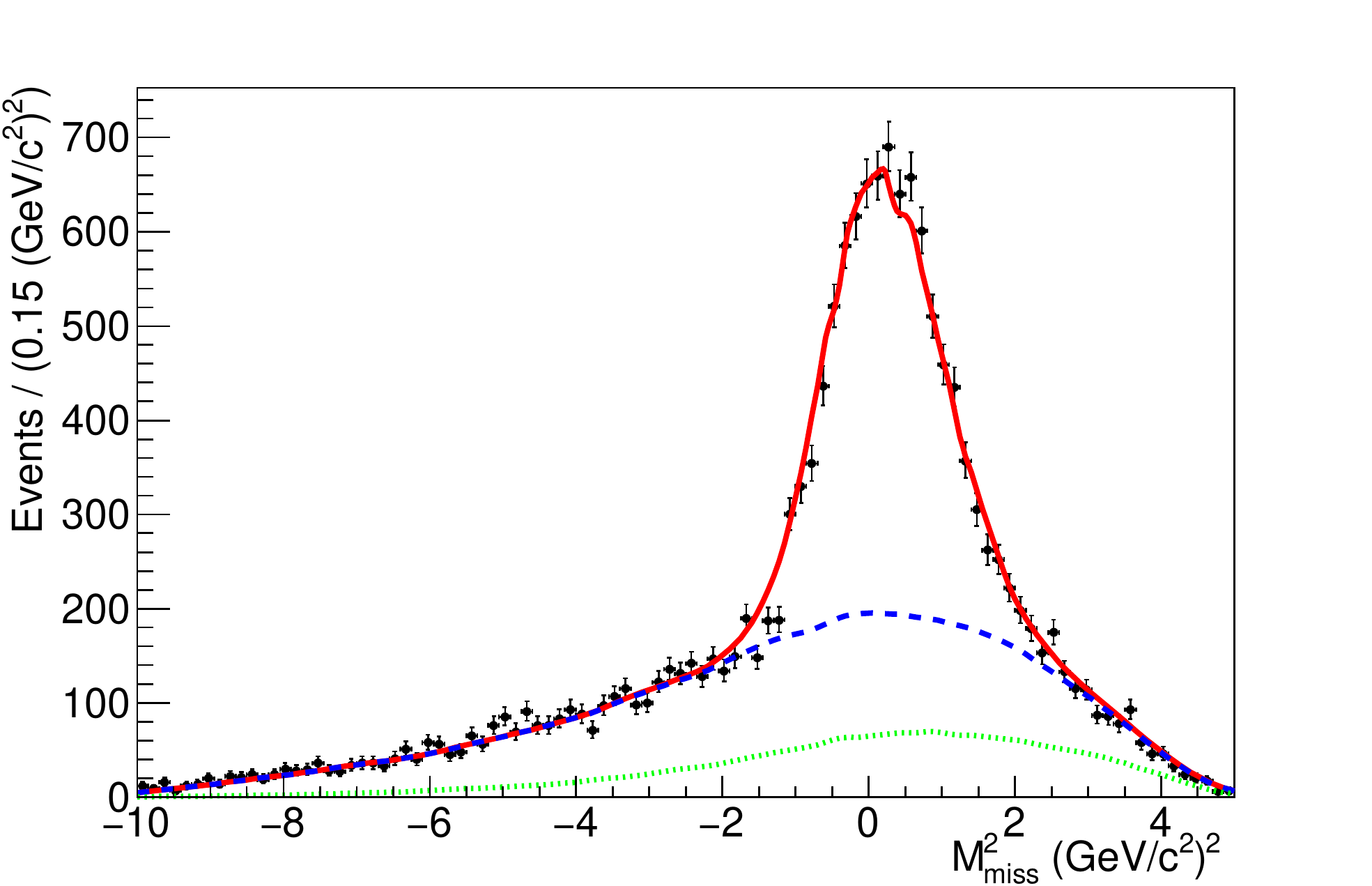}
  \includegraphics[width=7cm]{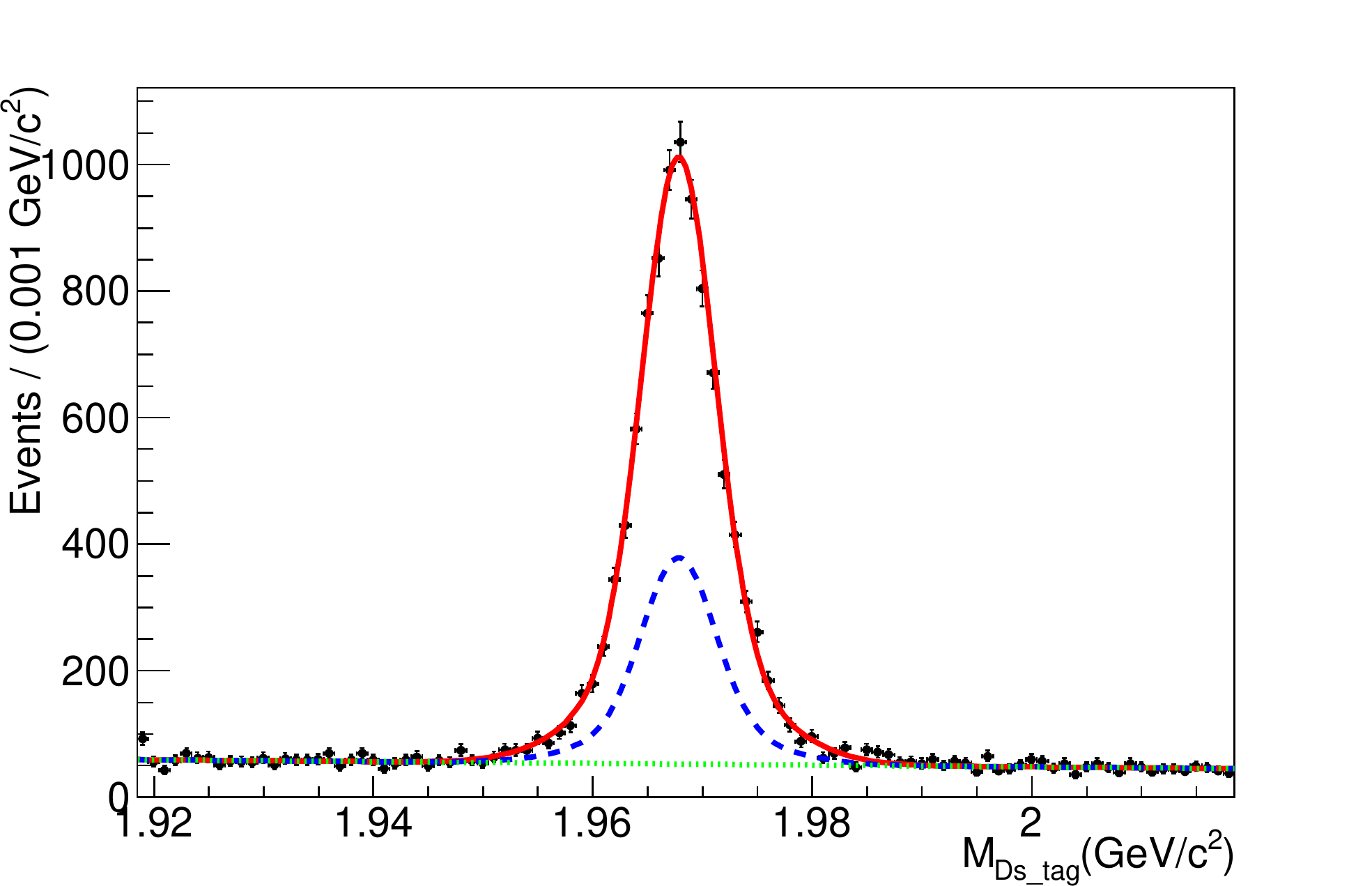}
  \caption{Distributions in $M_{\rm miss}^2$ (top) and  $D_s$ candidate mass (bottom) for tag candidates with $D_s\to\phi \pi$ in data (points with error bars), overlaid with fit results (cumulative): correct tags (red, solid),  incorrect tags with real $D_s$ (blue, dashed), and other incorrect tags (green, dotted).
In each plot, a signal band requirement is made on the quantity that is not displayed ($m_{D_s}^{PDG} \pm 0.015\ {\rm GeV}/c^2$, $|M_{\rm miss}^2|<2\ ({\rm GeV}/c^2)^2)$). 
  }
  \label{fig:tag-fit-data-all}
\end{figure}

The number of $B_s$ tags for each $D_s$ decay channel is found by a  binned {\color{black} 2D} maximum-likelihood fit of the distribution in $ M_{\rm miss}^2$ and the invariant mass of the $D_s$ candidate, $M_{D_{s\_{\rm tag}}}$, to a sum of three components, according to candidate origin:
\begin{enumerate}
  \item Correctly tagged candidates
  \item Incorrect tag, where a lepton from a $B_s$ semileptonic decay is paired with a real $D_s$ from the other $B_s$.  This can happen if $B_s$ mixing has occurred.
  \item Other incorrect tags: all other sources of candidates.  
  {\color{black}
  In addition to $B_s^{(*)}\bar B_s^{(*)}$ events, sources include $u\bar u$, $d\bar d$, $s \bar s$, $c\bar c$, and $B^{(*)}\bar B^{(*)}X$ events.}
\end{enumerate}
For each component, the $M_{\rm miss}^2$ distribution is taken to be a histogram obtained via Monte Carlo (MC) simulation.
For correctly reconstructed $D_s$, the distribution in $M_{D_{s\_{\rm tag}}}$ is represented by a sum of two Gaussians with a common mean. 
The widths of the Gaussians and their relative areas are obtained from MC simulation. 
For combinatorial $D_s$ background, each distribution is well-represented by a linear function. 
Tag decays, $B_s\to D_s X\ell\nu$, are modeled as a sum of $B_s \rightarrow D_s\ell \nu$ and $B_s \rightarrow D_s^*\ell \nu$;
all semileptonic $B_s$ decays to higher excited $D_s$ states observed to date involve $DK$ rather than $D_s$ in the final state, and decays including the states $D^*_{s0}(2317)$ and $D_{s1}(2460)$, which are known to decay to $D_s$, have not been observed\cite{pdg2020}.
The presence of higher mass excited $D_s$ in $D_s X \ell\nu$ final states would be manifested as a knee or bump to the right side of the $M_{\rm miss}^2$ peak.
The data are found to be consistent with contributions from $D_s$ and $D_s^*$ only (Figure \ref{fig:tag-fit-data-all}, top).
We find $N_{\rm tag}^{\phi \pi} =  6473 \pm 119$ and $N_{\rm tag}^{K_S^0 K} = 4435 \pm 126$.
The fit results for $D_s\to\phi\pi$  are shown in Fig.~\ref{fig:tag-fit-data-all}.

After selecting a $B_s$ candidate as the tag, we reconstruct the ``signal-side'' $D_s$ from the remaining tracks in the event. 
Candidates are reconstructed in all three modes discussed earlier, and we allow none of the tracks from the selected tag candidate to be used.
{\color{black}
The rate of signal $D_s$ in tagged events is determined through a binned 3D maximum-likelihood fit in the tag-side variables, $M_{\rm miss}^2$ and $M_{D_{s\_{\rm tag}}}$, and the invariant mass of the signal-side $D_s$ candidate, $M_{D_{s\_{\rm sig}}}$.
Each tag+signal candidate corresponds on the tag side to one of the three components comprising the 2D fit and on the signal side with a real or combinatorial $D_s$.
Events containing $B_s\to D_s X \ell\nu$ and inclusive $B_s \to D_s X$ may have a correctly reconstructed tag (component 1) with a signal $D_s$ or an incorrect tag (component 2) with a  $D_s$ that is actually from the tag side.  
We define the first type of event as ``signal'' and the second as ``cross-feed.''
Both types are included in our fit and used to determine the rate of $B_s \to D_s X$.

For signal events, where the tag-side (signal-side) $D_s$ decays to channel $i$ ($j$), the raw branching fraction (${\mathcal B}_{\rm raw}$) is found by dividing the number observed ($N_{{\rm sig};ij}$) by the total number of reconstructed tags in channel $i$ ($N_{{\rm tag};i}$), the branching fraction for the channel $j$ (${\mathcal B}_j$), and the reconstruction efficiency (${\cal E}_{ij;{\rm tag}}$) for $D_s$ in channel $j$: 
\begin{equation}
{\mathcal B}_{\rm raw}=\frac{N_{{\rm sig};ij}}{N_{{\rm tag};i}{\mathcal B}_j{\cal E}_{ij;{\rm tag}}}.
\end{equation}
We evaluate ${\cal E}_{ij;{\rm tag}}$ via MC for each pair of channels (Table~\ref{tab:eff-sig-ds}). 

For cross-feed events, the raw branching fraction is obtained through the relationship of their rate to that of signal events.
For both signal and cross-feed, the number of events found in a data set depends on many of the same factors: number of $B_s$ events, branching fractions of the reconstructed $D_s$ modes, branching fractions of $B_s \rightarrow D_sX \ell \nu$, and $B_s \rightarrow D_s X$.
The reason for this is clear: the two types have a common origin, differing only in the assigning of $D_s$ to tag- {\it vs} signal-side.
The differences stem from the selection processes and the fact that cross-feed is sourced only from the 50\% of events where $B_s \leftrightarrow \bar B_s$ mixing has occurred.
Thus, the expected ratio, $R_{ij}$ of observed cross-feed to signal events for each pair of $D_s$ decay channels is the ratio of selection efficiencies times 0.5.
These ratios are obtained via MC simulation.
From the number of observed cross-feed ($N_{{\rm cf};ij}$) we then have 
\begin{equation}
{\mathcal B}_{\rm raw}=\frac{N_{{\rm cf};ij}}{N_{{\rm tag};i}{\mathcal B}_j{\cal E}_{ij;{\rm tag}}R_{ij}}.
\end{equation}

}

A fit for ${\mathcal B}_{\rm raw}$ is performed simultaneously for the six $D_s$ tag-signal channel combinations, using the efficiencies and efficiency ratios determined as described above.
Intermediate branching fractions are fixed to PDG values. 

\begin{figure}[ht]
  \centering
  \includegraphics[width=5.6cm]{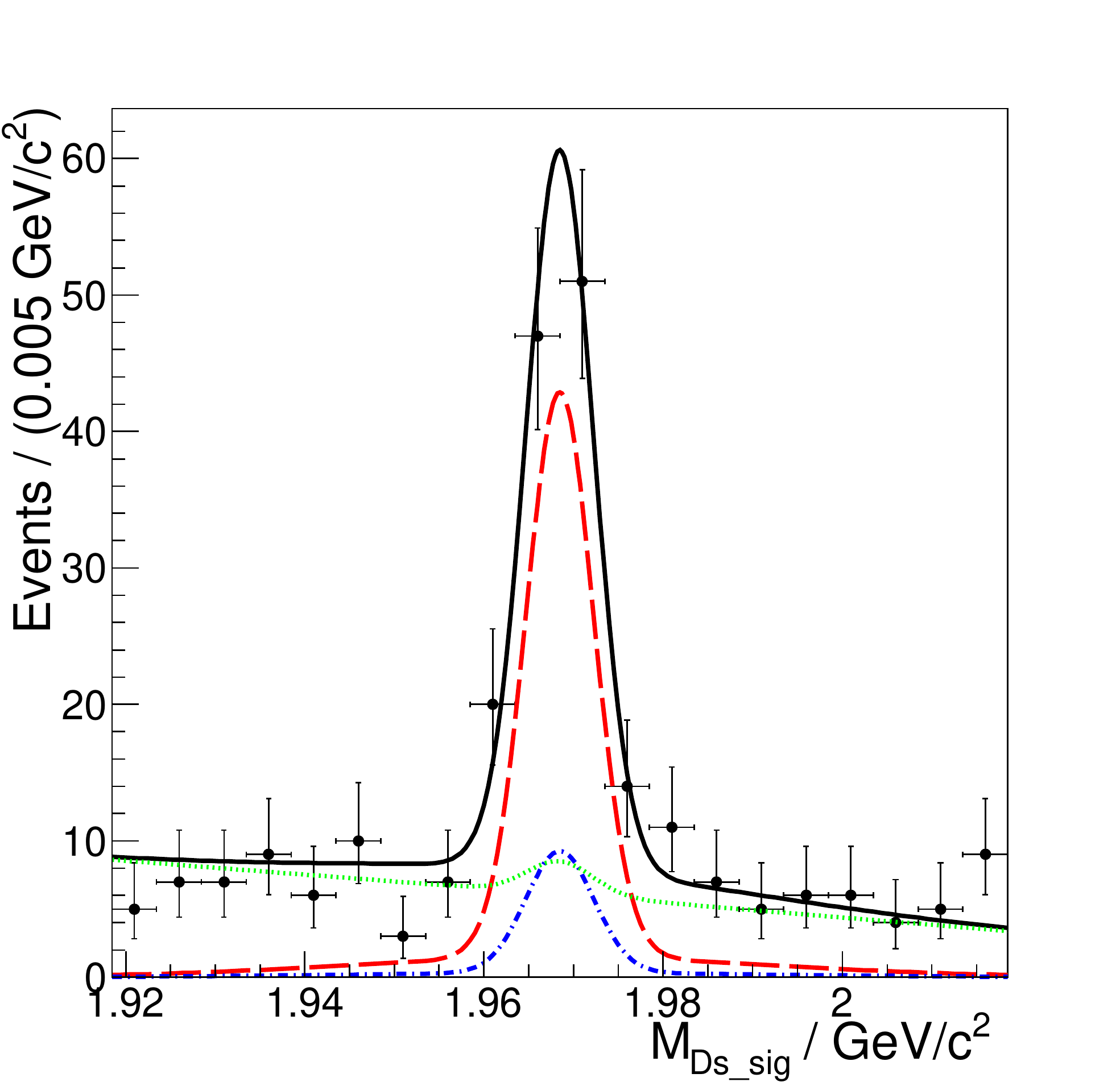}
  \includegraphics[width=5.6cm]{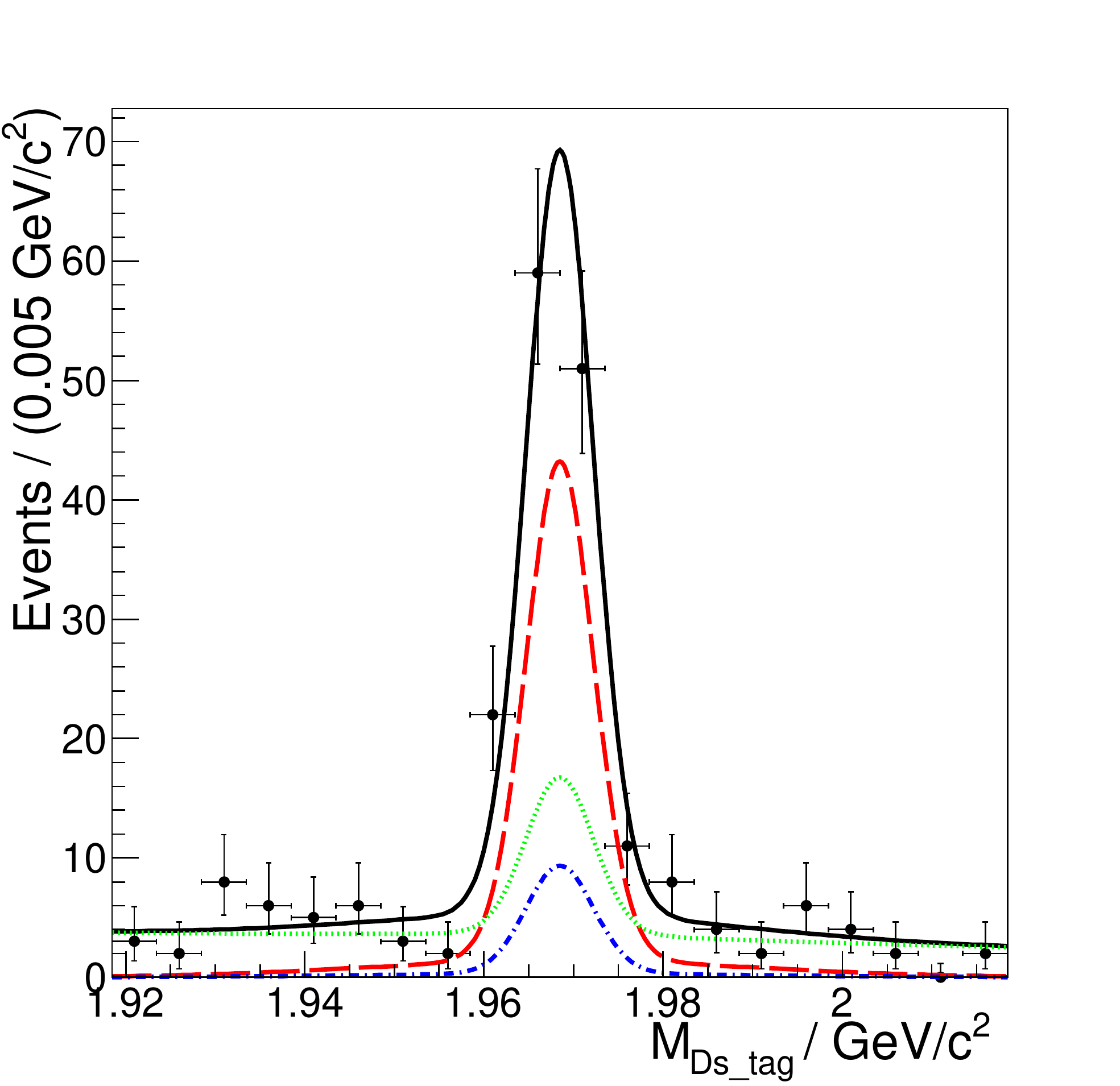}
  \includegraphics[width=5.6cm]{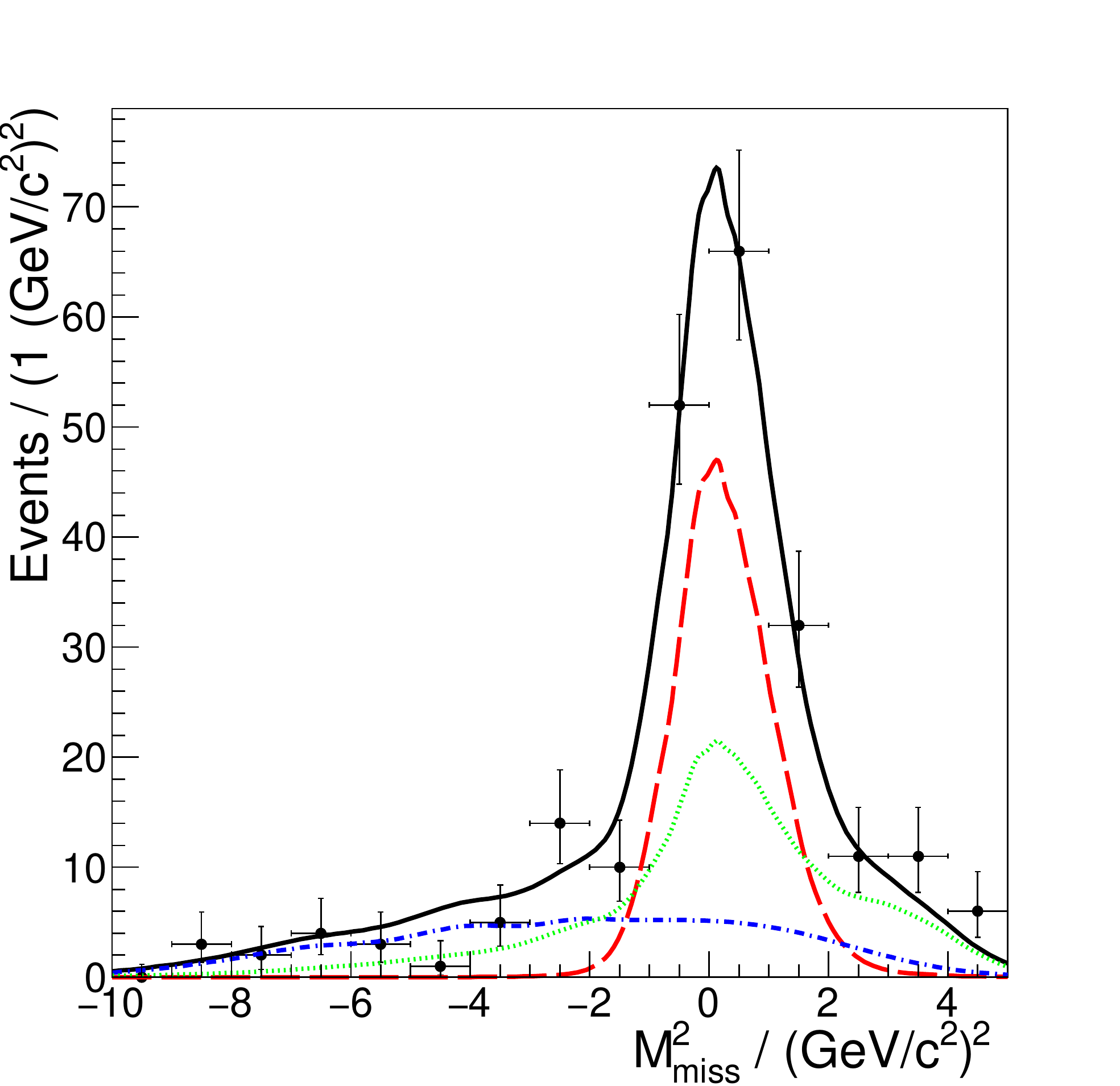}
  \caption{1D Projections of results from 3D fits, all $D_s$ modes combined, for $M_{Ds}^{\rm sig}$ (top), $M_{Ds}^{\rm tag }$ (center) and $M_{\rm miss}^2$ (bottom):
  data (points with error bars), signal (red, dashed), cross-feed (blue, dash-dotted), background (green, dotted), and total (black, solid).
For each projected variable, signal band requirements are made in the other two: 
  $M_{Ds}^{\rm sig}, M_{Ds}^{\rm tag} \in m_{Ds}^{\rm PDG} \pm 0.02 {\rm GeV}/c^2$,
  $M_{\rm miss}^2 \in [-2,2] ({\rm GeV}/c^2)^2$. 
  \label{fig:proj-data-all}
  }
\end{figure}

Our fit yields  ${\mathcal B}_{\rm raw} = (58.2 \pm 5.8)$\%, which corresponds to a fitted total of { $101 \pm 10$ signal and $36 \pm 4$} cross-feed events.
Projections of the fit are shown in Fig.~\ref{fig:proj-data-all}. 
To obtain ${\mathcal B}(B_s \rightarrow D_s X)$, we must make a correction to ${\mathcal B}_{\rm raw}$, due to the fact that the signal mode, $B_s \rightarrow D_s X$, is inclusive of the tagging mode, $B_s \rightarrow D_sX \ell \nu$. 
{\color{black}
We define ${\mathcal B}(B_s \rightarrow D_sX e \nu)+{\mathcal B}(B_s \rightarrow D_sX \mu \nu)\equiv {\mathcal B}_{ D_s\ell}$, ${\mathcal B}(B_s \rightarrow D_s X)\equiv {\mathcal B}_{D_s}$, and the respective reconstruction efficiencies $\epsilon_{ D_s\ell}$ and $\epsilon_{ D_s}$.
We take $N_{B_sB_s}$ to be the number of $B_s^{(*)}\bar B_s^{(*)}$ events.
The numbers of tags and signal are then
\begin{eqnarray}
N_{\rm tag} & = & N_{B_sB_s}( 2\epsilon_{ D_s\ell}{\mathcal B}_{ D_s\ell}-(\epsilon_{ D_s\ell}{\mathcal B}_{ D_s\ell})^2)\nonumber\\
 & = & N_{B_sB_s}\epsilon_{ D_s\ell}{\mathcal B}_{ D_s\ell}( 2-\epsilon_{ D_s\ell}{\mathcal B}_{ D_s\ell}),\\
N_{\rm sig} &=& N_{B_sB_s}(2\epsilon_{ D_s\ell}{\mathcal B}_{ D_s\ell}\epsilon_{ D_s}{\mathcal B}_{ D_s}-(\epsilon_{ D_s\ell}{\mathcal B}_{ D_s\ell})^2)\nonumber\\
&=& N_{B_sB_s}\epsilon_{ D_s\ell}{\mathcal B}_{ D_s\ell}(2\epsilon_{ D_s}{\mathcal B}_{ D_s}-\epsilon_{ D_s\ell}{\mathcal B}_{ D_s\ell}).
\end{eqnarray}

Their ratio, corrected for efficiencies, is ${\mathcal B}_{\rm raw}$: 
\begin{eqnarray}
{\mathcal B}_{\rm raw} &=& \frac{N_{\rm sig}/\epsilon_{ D_s}}{N_{\rm tag}}\nonumber\\
&=&\frac{2\epsilon_{ D_s}{\mathcal B}_{ D_s}-\epsilon_{ D_s\ell}{\mathcal B}_{ D_s\ell}}{\epsilon_{ D_s}(2-\epsilon_{ D_s\ell}{\mathcal B}_{ D_s\ell})}\nonumber\\
&=&\frac{{\mathcal B}_{ D_s}-\frac{\epsilon_{ D_s\ell}}{2\epsilon_{ D_s}}{\mathcal B}_{ D_s\ell}}{1-\epsilon_{ D_s\ell}{\mathcal B}_{ D_s\ell}/2}.
\end{eqnarray}
Thus,
\begin{eqnarray}
  \label{eq:corr}
{\mathcal B}_{D_s}
  &=& {\mathcal B}_{\rm raw}\left(1-\frac{\epsilon_{ D_s\ell}{\mathcal B}_{ D_s\ell}}{2}\right) +\frac{\epsilon_{ D_s\ell}}{2\epsilon_{ D_s}}{\mathcal B}_{ D_s\ell}.
\end{eqnarray}
To estimate $\epsilon_{ D_s\ell}$, we use  ${\mathcal B}_{D_s\ell}$= $(16.2 \pm 2.6({\rm sys})) \%$~\cite{pdg2020}, $2N_{B_sB_s}=(1.66\pm 0.27({\rm sys}))\times 10^7$\cite{Esen}, $N_{\rm tag}=10908\pm 173 ({\rm stat})$ (our measurement), where errors are indicated as being statistical or systematic in origin. 
We calculate $\epsilon_{ D_s}$ from Table~\ref{tab:eff-sig-ds} and branching fractions from~\cite{pdg2020}.
We find
\begin{eqnarray}
\epsilon_{ D_s\ell} &\approx& \frac{N_{\rm tag}}{2N_{B_sB_s}{\mathcal B}_{ D_s\ell}}\nonumber\\
&=&(4.1\pm 0.1({\rm stat})\pm0.7({\rm sys}))\times 10^{-3},\\
\epsilon_{ D_s} &=& \sum_i \epsilon_i {\cal B}_i \nonumber\\
&=&(1.62\pm 0.03({\rm sys}))\times 10^{-2}.
\end{eqnarray}

The first correction term is found to be negligible (less than $10^{-3}$), and the second is
\begin{eqnarray}
  \label{eq:corr}
{\mathcal B}_{D_s}-{\mathcal B}_{\rm raw}
  &=&  \frac{\epsilon_{ D_s\ell}}{2\epsilon_{ D_s}}{\mathcal B}_{ D_s\ell}\nonumber\\
  &=& (2.03\pm 0.03({\rm stat})\pm 0.33({\rm sys}))\nonumber\\
  && \times 10^{-2}.
\end{eqnarray}
We thus find
\begin{eqnarray}
{\mathcal B}_{D_s}  &=& (60.2\pm 5.8\pm 0.3)\%.
\end{eqnarray}
As a cross-check of our method, we fit for signal while floating the cross-feed component and find  ${\mathcal B}_{\rm raw} = (64.8 \pm 8.1)$\%, which is consistent with our result.

{\color{black}

To confirm the 3D fitting procedure and correction to ${\mathcal B}_{\rm raw}$, we generated ensembles of simulated data distributions with varied signal content.
Signal and crossfeed distributions were generated by randomly selecting from our large sample of MC-generated signal events.
For background we generated distributions according to those used in the fit, with parameters fixed to the results of the fit to data.
Ensembles of 200 experiments were generated for each of ten branching fractions in the range 10-100\%, in 10\% increments.
Each distribution was fitted according to our procedure.
The resulting ensemble mean branching fractions, corrected and plotted against input branching fractions, were fitted to a line.
This test was repeated for each of the six $D_s$ mode combinations, as well as the combined set.
All showed consistency with a unit slope and no systematic shifts.
}

}

Our estimates of systematic uncertainties are summarized in Table~\ref{tab:sys-summary}.
We evaluate the effects from the considered sources by varying each and taking the resulting shift observed in ${\mathcal B}_{\rm raw}$ as the uncertainty.
In cases affecting the $D_s$ mode combinations separately, the maximum excursion is taken as a conservative estimate of the uncertainty on the combined result.
Because this measurement involves tagging, many of the systematic uncertainties associated with tagging cancel approximately in taking the ratio of tags, with and without signal.
The effect from the uncertainty due to the composition and model of $B_s \rightarrow D_s X\ell \nu$ on efficiency and on the $M_{\rm miss}^2$ fitting shape is estimated by varying the relative rates of $B_s \rightarrow D_s\ell \nu$ and $B_s \rightarrow D_s^*\ell \nu$ within the uncertainties\cite{pdg2020} and by varying the HQET2 parameters in the MC generator by $\pm$10\%.
For the ``other incorrect tag'' (type 3, above), the $M_{\rm miss}^2$ distribution in data from tags with ``sideband'' $D_s$ candidates, $|M_{\rm cand} - m_{D_s}\pm 40| < 10$ MeV, is substituted in the fit. 
Uncertainties due to fitting of the $D_s$ mass distributions are determined by changing the signal shape from two Gaussians to three and the background from a first-order to a second-order polynomial.
We vary each ratio of signal to cross-feed efficiency in the fit by $\pm 1\sigma$.
The uncertainties due to branching fractions of the reconstructed $D_s$ decays are estimated by varying each by $\pm$1$\sigma$\cite{pdg2020} of its value in the fitting procedure.
The reconstruction efficiencies are varied by the amount of their statistical error from the MC sample.
The uncertainty due to the limited statistical power of our linearity test is estimated by varying the parameters from the linear fit by $\pm 1\sigma$.
To estimate effects from our selection of a single tag candidate per event, we reanalyze the data using random selection and take the shift in the result to be the uncertainty.

The uncertainty on the tracking efficiency affects only the three signal-side tracks comprising the $D_s$ candidate and is estimated to be 0.35\% per track, thus, we take 1.1\% as the uncertainty from this source.
The systematic uncertainty from $K$-$\pi$ identification efficiencies is estimated to be 1.3\%.

The fitted shape of the $M_{\rm miss}^2$ distribution depends on the  $B_s^*-B_s$ mass difference, $\delta E/c^2$, and its uncertainty may affect the fit in two ways: in the value used to generate the MC signal events ({\it vs} the actual value) and in the value used to calculate $M_{\rm miss}^2$.
For this analysis, the values are $45.9$~MeV$/c^2$ for MC generation and $47.3$~MeV$/c^2$ for $M_{\rm miss}^2$. 
The PDG presents two numbers, $(46.1\pm 1.5)$~MeV$/c^2$ as a world average and a PDG fit of $(48.6^{+1.8}_{- 1.5})$~MeV$/c^2$ \cite{pdg2020}.
As $M_{\rm miss}^2$ is fitted in both the numerator and denominator to obtain ${\mathcal B}_{\rm raw}$, effects from such differences are expected to cancel, at least in part. 
To estimate possible systematic shifts due to these differences, we vary separately the calculation using $\delta E/c^2$ and the value used in MC generation in the range 45.9-49.0~MeV/$c^2$.
Changing the calculation of $M_{\rm miss}^2$ results in a maximum excursion in ${\mathcal B}_{\rm raw}$ of less than 0.1\%.
Changing the value in the MC generator results in a maximum excursion of 1.2\%. 
We assign an uncertainty of 1.2\%.

\begin{figure}
  \centering
  \includegraphics[width=7cm]{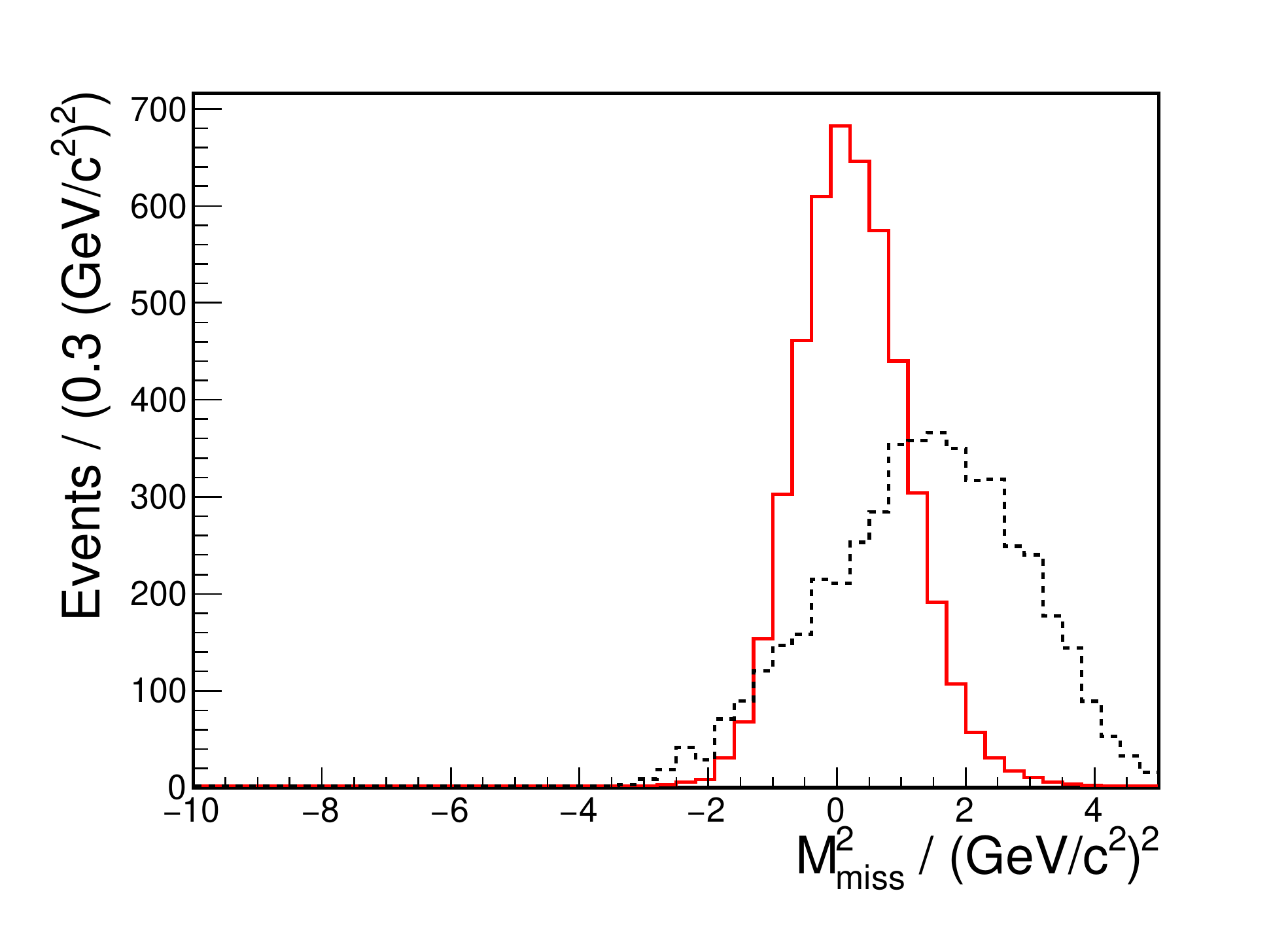}
  \caption{The distributions in $M_{\rm miss}^2$ for  $B_s~\rightarrow~D_s X \ell \nu$ (red) and $B \rightarrow D_s K X \ell \nu$ (black), with $D_s\to \phi\pi$.
}
  \label{fig:sys-mm2-b-tag}
\end{figure}

We consider possible contributions to the tag-side sample from the non-strange $B$ decay ${\mathcal B}(B \rightarrow D_s^{(*)} K \ell \nu)$, which is not included in our generic MC generator. 
We use ${\mathcal B}(B^+\rightarrow D_s^{(*)-} K^+ \ell^+ \nu)=(6.1\pm 1.0)\times 10^{-4}$\cite{pdg2020}, assume that ${\mathcal B}(B^0\rightarrow D_s^{(*)-} K^0 \ell^+ \nu)$ is the same, and multiply by a factor of two to account for both electrons and muons.
Taking ${\mathcal B}(\Upsilon(5{\rm S})\rightarrow B\bar{B}X)=$76\%, ${\mathcal B}(\Upsilon(5{\rm S})\rightarrow B_s\bar{B}_sX)=$20\%, and ${\mathcal B}(B_s \rightarrow X \ell \nu)=9.6\%$\cite{pdg2020}, 
we estimate 
\begin{equation}
  \frac{{\mathcal B}(\Upsilon(5{\rm S}) \rightarrow B\bar{B}X) \cdot {\mathcal B}(B \rightarrow D_s^{(*)} K \ell \nu)} 
  {{\mathcal B}(\Upsilon(5{\rm S}) \rightarrow B_s^{(*)}\bar B_s^{(*)}) \cdot {\mathcal B}(B_s \rightarrow D_s X \ell \nu)}
  \approx 0.048.
\end{equation}
As the shape in  $M_{\rm miss}^2$ includes a kaon in addition to the neutrino, it is expected to peak more broadly and at a higher value than does the $B_s$ channel.
This is confirmed in studies of MC-generated $B\bar B X$ events containing $B \rightarrow D_s^{(*)} K \ell \nu$ in the $D_s$ tag modes.
Fig.~\ref{fig:sys-mm2-b-tag} illustrates the difference.
We measure the effect on our MC tag fit of including such events, and estimate a  contribution to ${\mathcal B}(B_s \rightarrow D_s X \ell \nu)$ of  $< 0.02$\% (0.5\%) to the $D_s\to\phi\pi$ ($D_s\to K_S^0 K$) channel.
We assign an overall systematic uncertainty of 0.5\%.
The uncertainties from the above sources are summed in quadrature to arrive at the total fractional systematic uncertainty in ${\mathcal B}_{\rm raw}$ of 3.8\%.
\begin{table*}[tb]  
  \centering
  \caption{Systematic uncertainties on ${\mathcal B}_{\rm raw}$, in \%. 
  The total is the   sum in quadrature from all sources.}
  \label{tab:sys-summary}
  \begin{tabular}{ lccccccc}
    \hline \hline
    \multirow{3}{*}{Source} & \multicolumn{6}{c}{Channel} & 
    \multirow{3}{*}{Combined} \\
    & \multicolumn{3}{c}{$\phi \pi$ Tag} & \multicolumn{3}{c}{$K_S^0
    K$ Tag} & \\
    & $\phi \pi$ & $K_S^0 K$ & $K^{*0} K$ & $\phi \pi$ & $K_S^0 K$ &
    $K^{*0} K$ & \\
    \hline
    Model, tag & \multicolumn{3}{c}{1.5} & \multicolumn{3}{c}{1.1} &    1.5 \\
    Model, signal & 0.1 & 0.1 & 0.3 & 0.1 & 0.1 & 0.1 
    & 0.3 \\
    Model, cross-feed & 0.4 & 0.3 & 0.3 & 0.2 & 0.1 &
    0.1 & 0.4 \\
    $M_{\rm miss}^2$ shape, $M_{B_s^*}-M_{B_s}$ & \multicolumn{3}{c}{1.2} & \multicolumn{3}{c}{1.2} &    1.2 \\
    $M_{\rm miss}^2$ background & 0.1 & 0.2 & 0.1 & 0.5 & 0.2 &
    0.3 & 0.5 \\
    $M(D_s)$ signal shape & 0.2 & 0.2 & 1.2 & 0.1 & 0.1 & 
    1.0
    & 1.2 \\
    $M(D_s)$ background shape & 1.0 & 0.6 & $<$0.1 & $<$0.1 & 0.1 &
    0.1 & 1.0 \\
    Cross-feed efficiency & 0.5 & 0.3 & 0.6 & 0.3 & 0.1 & 0.3 & 0.6 \\
    Reconstruction efficiency  & 0.4 & 0.2 & 0.4 & 0.2 & 0.1 & 0.2 & 0.4 \\
    Statistics, linearity test & 0.2 & 0.3 & 0.3 & 0.3 & 0.4 & 0.4 & 0.4\\
    $B\to D_s^{(*)}K\ell\nu$  & \multicolumn{3}{c}{$<$0.02}& \multicolumn{3}{c}{0.5} & 0.5 \\
    ${\cal B}(D_s\to\phi\pi)$  & \multicolumn{6}{c}{-} & 1.2 \\
    ${\cal B}(D_s\to K_S^0K)$  & \multicolumn{6}{c}{-} & 0.5 \\
    ${\cal B}(D_s\to K^{*0}K)$  & \multicolumn{6}{c}{-} & 1.2 \\
    Single tag selection & \multicolumn{6}{c}{-} & 1.0 \\
    Tracking & \multicolumn{6}{c}{-} & 1.1 \\
    K-$\pi$ identification & \multicolumn{6}{c}{-} & 1.3 \\
    \hline
    Total & \multicolumn{6}{c}{-} & 3.8 \\
    \hline \hline
  \end{tabular}
\end{table*}
Adding the systematic uncertainties in quadrature, we find 
\begin{eqnarray}
{\mathcal B}(B_s \rightarrow D_s X) = (60.2 \pm 5.8 \pm 2.3)\%. 
\end{eqnarray}
The central value is lower than the theoretical expectation ($86^{+8}_{-13}$)\%~\cite{suzuki:1985}, and $\approx 1.3\sigma$ below the world average, (93 $\pm$ 25)\%\cite{pdg2020}.
Given the history of uncertainty on the rates and composition of charm states at higher mass in $B$ decay, a lower value may be explained by a rate of $c\bar s$ to $D$ {\it vs.} $D_s$ that is higher than anticipated.
The implications of a lower central value are notable.
Experimentally, the value affects the derived fraction $f_s$ of $B_s$ events among $\Upsilon$(5S) decays, which impacts the absolute normalization of all $B_s$ branching fractions measured via $\Upsilon$(5S) decays.
In the earlier Belle measurements of $f_s$\cite{belle-result,Esen}, Eq.~\ref{eq:Bs2Dx} was used with $f_q= 1 - f_s$. 
More recently, it has been found that there is a nonzero rate to bottomonia, including $\Upsilon(1S)$, $\Upsilon(2S)$, $\Upsilon(3S)$, $h_b(1P)$ and $h_b(2P)$.
We take the rate of events with ``no open bottom'' to be 
$f_{\rm nob}=4.9^{+5.0}_{-0.6}$\%\cite{Mizuk}. 
Charm is highly suppressed in these decays, so we take $f_q = 1 - f_s - f_{\rm nob}$. 
Using ${\mathcal B}(\Upsilon(5{\rm S}) \rightarrow D_s X)$= 
($45.4 \pm 3.0$)\%\cite{DrutskoyUpdate} 
and  ${\mathcal B}(B \rightarrow D_s X)$=($8.3 \pm 0.8$)\% \cite{pdg2020}, we solve Eq.~\ref{eq:Bs2Dx} for $f_s$ and find
\begin{equation}
f_s = 0.285 \pm 0.032 ({\rm stat}) \pm 0.037 ({\rm sys}).
\label{eq:fscalc}
\end{equation}
This value is larger than the world average, $f_s=0.201 \pm
0.031$\cite{pdg2020}, which is evaluated assuming the model-based estimates ${\mathcal B}(B_s \rightarrow D_s X)=(92\pm 11)\%$ and ${\mathcal B}(B_s \rightarrow D^0 X)=(8\pm 7)\%$\cite{belle-result}; 
the impact of introducing $f_{\rm nob}$ to the calculation is minor.
Our result uses the same value of ${\mathcal B}(\Upsilon(5{\rm S}) \rightarrow D_s X)$ from which $f_s$ is derived in~\cite{Esen} and thus supersedes the value presented there.
It is consistent with a recent Belle measurement of $f_s$ by an independent method\cite{Mizuk}.
An older Belle measurement of $f_s$ from semileptonic decays\cite{Oswald} assumed that only $D_{s1}$ and $D_{s2}$ contribute to non-strange charm, $B_s\to DKX\ell\nu$.
Given recently reported evidence of substantial contributions from nonresonant $DK(X)$\cite{LHCb}, this value is likely an underestimate, so we do not compare it with the result reported here.

Applying Eq.~\ref{eq:Bs2Dx} with ${\mathcal B}(B \rightarrow D^0/\bar D^0 X)$=($61.5 \pm 2.9$)\%\cite{pdg2020}, ${\mathcal B}(\Upsilon(5{\rm S})\to D^0 X)=(108 \pm 8$)\% \cite{pdg2020} , and our result for $f_s$, we find
${\mathcal B}(B_s \rightarrow D^0 X)$=($46 \pm 2 ({\rm stat}) \pm 20 ({\rm sys})$)\%, 
where the systematic uncertainties on ${\mathcal B}(\Upsilon(5{\rm S})\to D^0 X)$ and $f_{\rm nob}$ dominate.
This value is consistent with our finding of a lower rate of $D_s$ from $B_s$ decay, as the total charm content would need to be accounted for by an increased rate of nonstrange charm.
No experimental results for $B_s \rightarrow D^0 X$ are currently included in the PDG tables\cite{pdg2020}.

To summarize, we have made the first direct measurement of the $B_s \rightarrow D_s X$ inclusive branching fraction, using a $B_s$ semileptonic tagging method at the $\Upsilon$(5S) resonance. 
We find
\begin{eqnarray}
{\mathcal B}(B_s \rightarrow D_s X) = (60.2 \pm 5.8({\rm stat}) \pm 2.3({\rm sys}))\%, 
\end{eqnarray}
which is substantially lower than the world average but consistent within its large uncertainties.
This result is used to recalculate the fraction $f_s$ of  $\Upsilon$(5S) events containing $B_s$, 
\begin{eqnarray}
f_s = 0.285 \pm 0.032 ({\rm stat}) \pm 0.037 ({\rm sys}).
\end{eqnarray}
This value supersedes that reported in \cite{Esen}.


We thank the KEKB group for the excellent operation of the
accelerator; the KEK cryogenics group for the efficient
operation of the solenoid; and the KEK computer group, and the Pacific Northwest National
Laboratory (PNNL) Environmental Molecular Sciences Laboratory (EMSL)
computing group for strong computing support; and the National
Institute of Informatics, and Science Information NETwork 5 (SINET5) for
valuable network support.  We acknowledge support from
the Ministry of Education, Culture, Sports, Science, and
Technology (MEXT) of Japan, the Japan Society for the 
Promotion of Science (JSPS), and the Tau-Lepton Physics 
Research Center of Nagoya University; 
the Australian Research Council including grants
DP180102629, 
DP170102389, 
DP170102204, 
DP150103061, 
FT130100303; 
Austrian Federal Ministry of Education, Science and Research (FWF) and
FWF Austrian Science Fund No.~P~31361-N36;
the National Natural Science Foundation of China under Contracts
No.~11435013,  
No.~11475187,  
No.~11521505,  
No.~11575017,  
No.~11675166,  
No.~11705209;  
Key Research Program of Frontier Sciences, Chinese Academy of Sciences (CAS), Grant No.~QYZDJ-SSW-SLH011; 
the  CAS Center for Excellence in Particle Physics (CCEPP); 
the Shanghai Pujiang Program under Grant No.~18PJ1401000;  
the Shanghai Science and Technology Committee (STCSM) under Grant No.~19ZR1403000; 
the Ministry of Education, Youth and Sports of the Czech
Republic under Contract No.~LTT17020;
Horizon 2020 ERC Advanced Grant No.~884719 and ERC Starting Grant No.~947006 ``InterLeptons'' (European Union);
the Carl Zeiss Foundation, the Deutsche Forschungsgemeinschaft, the
Excellence Cluster Universe, and the VolkswagenStiftung;
the Department of Atomic Energy (Project Identification No. RTI 4002) and the Department of Science and Technology of India; 
the Istituto Nazionale di Fisica Nucleare of Italy; 
National Research Foundation (NRF) of Korea Grant
Nos.~2016R1\-D1A1B\-01010135, 2016R1\-D1A1B\-02012900, 2018R1\-A2B\-3003643,
2018R1\-A6A1A\-06024970, 
2019K1\-A3A7A\-09033840,
2019R1\-I1A3A\-01058933
2021R1\-A6A1A\-03043957,
2021R1\-F1A\-1060423, 
2021R1\-F1A\-1064008;
Radiation Science Research Institute, Foreign Large-size Research Facility Application Supporting project, the Global Science Experimental Data Hub Center of the Korea Institute of Science and Technology Information and KREONET/GLORIAD;
the Polish Ministry of Science and Higher Education and 
the National Science Center;
the Ministry of Science and Higher Education of the Russian Federation, Agreement 14.W03.31.0026, 
and the HSE University Basic Research Program, Moscow; 
University of Tabuk research grants
S-1440-0321, S-0256-1438, and S-0280-1439 (Saudi Arabia);
the Slovenian Research Agency Grant Nos. J1-9124 and P1-0135;
Ikerbasque, Basque Foundation for Science, Spain;
the Swiss National Science Foundation; 
the Ministry of Education and the Ministry of Science and Technology of Taiwan;
and the United States Department of Energy and the National Science Foundation.


\begin{thebibliography}{99}
\bibitem{chargeconjugate}
Inclusion of charge-conjugate modes is implied throughout this paper.
\bibitem{DrutskoyUpdate}
  This value is obtained by Belle using 121.4~fb$^{-1}$ of data and the method described in Ref. \cite{belle-result}, with $D_s^+$ reconstructed in the mode $K^+K^-\pi^+$.
\end{thebibliography}
\end{document}